\documentclass[
 preprint,
 amsmath,amssymb,
 aps,
]{revtex4-1}

\usepackage{graphicx}
\usepackage{dcolumn}
\usepackage{bm}

\begin{document}


\title{Power-law citation distributions are not scale-free\\}

\author{Michael Golosovsky}
\email{michael.golosovsky@mail.huji.ac.il}

\affiliation{The Racah Institute of Physics, The Hebrew University of Jerusalem, 91904 Jerusalem, Israel\\
}%
\date{\today}

\begin{abstract}
We analyze time evolution of statistical distributions of citations to scientific papers published in one year. While these distributions  can be fitted by a power-law dependence we find that they are nonstationary and the exponent of the power law fit decreases with time and does not come to saturation. We attribute the nonstationarity of citation distributions to different longevity of the low-cited and highly-cited papers. By measuring citation trajectories of papers we found that citation careers of the low-cited papers come to saturation after 10-15 years while those of the highly-cited papers continue to increase  indefinitely: the papers that exceed some citation threshold become runaways. Thus, we show that although citation distribution can look as a power-law, it is not scale-free and there is a hidden dynamic scale associated with the onset of runaways. We compare our measurements to our recently developed model of citation dynamics based on copying/redirection/triadic closure and find explanations to our empirical observations.
\begin{description}
\item[PACS numbers]01.75.+m, 02.50.Ey, 89.75.Fb, 89.75.Hc
\end{description}
\end{abstract}
\maketitle
\section{Introduction}
Highly-skewed statistical distributions were discovered more than a century ago and up to now remain an object of intense  research (see Refs. \cite{Mitzenmacher2004,Newman2005,Clauset2009,Pinto2012} for  comprehensive reviews). The most important class of  highly-skewed continuous distributions is the power-law
\begin{equation}
p(x)\propto x^{-\alpha};\; x\geq x_{min}
\label{ParetoI}
\end{equation}
and the shifted power-law (Pareto II)
\begin{equation}
p(x)\propto(x+w)^{-\alpha};\; x\geq x_{min}
\label{ParetoII}
\end{equation}
where $p(x)$ is the probability density function, $\alpha$ is the exponent, and $w$ is the shift. In contrast to the Gaussian distribution with its finite moments, the moments of the power-law distributions  can diverge. In particular,  the mean diverges  for $\alpha\leq2$ and the variance diverges for $\alpha\leq3$.  Thus, the first and the second moments of a power-law distribution with $2\leq\alpha\leq3$ are determined by its tail and this is the reason why such distribution  is named heavy-tailed.

The discrete analogue of Eq. \ref{ParetoII} is frequently represented by the Waring distribution \cite{Glanzel2004,Burrell2005,Mingers2006} which is  also known as the shifted Yule-Simon distribution \cite{Mitzenmacher2004,Newman2005,Clauset2009},
\begin{equation}
p(k)=\frac{B(k+w,\alpha)}{B(w,\alpha-1)}.
\label{Waring}
\end{equation}
Here, $B(a,b)=\frac{\Gamma(a)\Gamma(b)}{\Gamma(a+b)}$ is the Euler beta-function. The parameter $\alpha$ is the analogue of the exponent and $w$  is the shift. The mean of the Waring distribution is
\begin{equation}
M=\frac{w}{\alpha-2}
\label{mean}
\end{equation}
and it obviously diverges for $\alpha\leq2$. For $k>>w$  Eq. \ref{Waring}  reduces to the Zipf-Mandelbrot distribution,
\begin{equation}
p(k)\approx(\alpha -1)\frac{w^{\alpha-1}}{(k+w)^{\alpha}}.
\label{Tsallis}
\end{equation}
which is a discrete analogue of Eq. \ref{ParetoII}.

Another class of the highly-skewed distributions is the log-normal,
\begin{equation}
p(x)dx=\frac{1}{\sigma\sqrt{2\pi}} e^{-\frac{(\ln x-\mu)^{2}}{2\sigma^{2}}}d\ln x,
\label{lognormal}
\end{equation}
where $\mu$ and $\sigma$ characterize, correspondingly, the mean and the variance.

A peculiar property of the power-law and log-normal distributions  is that they are scale-free, namely, the density functions $p(x)$  and $p(x/S)$, where $S$ is a constant, have the same shape and are just shifted on the log-log scale (this is also true  for Eq. \ref{Waring}  when $k>>w$), in other words, these distributions or at least their tails are self-similar. The scale-free property of the highly-skewed distributions  has been a source of fascination for many physicists that sought deep analogies with other scale-free phenomena such as fractals, phase transitions and critical phenomena.\cite{Dorogovtsev2001,Newman2005,Caldarelli2007,Willinger2009,Sornette2012,Barabasi2015}

Complex networks provide an  abundant source of  highly-skewed distributions.\cite{Seglen1992} The first object  identified as a complex network  was citations to scientific papers.\cite{SollaPrice1965} This occurred in 70-ies and the field of complex networks laid dormant until the appearance of Internet  and other information networks in 90-ies.  Thereafter complex networks came to forefront of the Physics and computer science research. Indeed, design of Internet browsers and search engines strongly relies on the degree distribution in the World Wide Web. This drew much attention to the characterization of the degree  distribution in  WWW and other complex networks as well.\cite{Adamic2000,Barabasi2015}  The results of diverse measurements  generated common belief that degree distributions in complex networks are described by the power-law dependence  with the exponent $2\leq\alpha\leq3$.

How  was this conclusion drawn from measurements?  The simplest way to characterize the degree distribution  is to plot it on the log-log scale. The straight line indicates the  power-law  while the parabola  suggests the log-normal distribution. In practice, the test for curvature on the log-log scale doesn't discriminate well between these two distributions. Indeed, if the log-normal distribution is very wide,  then a large piece of parabola looks like a straight line.   For discrete distributions the situation is even worse since the log-log plot of the discrete power-law distribution (Eq. \ref{Waring}) also has convex shape at small degrees and this further exacerbates the problem of distinguishing this distribution from the log-normal.

Even if the log-log plot of a statistical distribution or of a part of it looks like a straight line, to find its slope is not an easy task.\cite{Thelwall2016} Since most distributions round up at small degrees, to find the slope one shall cut the small degrees and focus on the tail of the distribution. This cutoff procedure is subjective and is a source of uncertainty.\cite{Clauset2009} The difficulties of experimental  identification of the power-law degree distribution  generated  a  substantial controversy of whether degree distributions in complex networks are better described by the power-law, log-normal, or stretched exponential.\cite{Mitzenmacher2004,Newman2005,Clauset2009}  The history of assessment of  citation distributions is a good example of such controversy.  Beginning from the works of de Solla Price \cite{SollaPrice1965} citations were fitted by a discrete power-law distribution with the exponent $\alpha=$ 2.5-3.16.\cite{Redner1998,Clauset2009} Subsequently, citations were claimed to follow log-normal or discretized log-normal distribution \cite{Limpert2001,Redner2005,Stringer2008,Evans2012,Thelwall2016} with $\sigma=1-1.2$. Recent encompassing studies \cite{Albarran2011,Brzezinski2015} claimed  again the power-law  distribution with the exponent  $\alpha$ varying between 3 and 4.

Why is it so important to find the functional form of degree distribution of a complex network? The "big" question  is how these complex networks grow, what is their generative mechanism. The motivation for precise characterization of the network degree  distribution  is  driven by the belief \cite{Mitzenmacher2005,Stumpf2012} that the functional form  of this distribution is a clue to the mechanism of network growth.

The most widely discussed network growth mechanism  is the cumulative advantage/ preferential attachment.\cite{Price1976,Barabasi2015} This is  an umbrella term unifying several closely related mechanisms that include ageing, fitness, and nonlinearity \cite{Dorogovtsev2000a,Krapivsky2001,Krapivsky2005,Newman2005,Bianconi2001,Wang2013}. Barabasi \cite{Barabasi2015} showed theoretically that if some complex network grows according to the linear preferential attachment rule  then, in the long time limit,  its degree distribution  converges to the power-law with almost universal  exponent $\alpha=3$.  Following this seminal study, the power-law degree distribution in  complex networks was considered as a proof of the preferential attachment growth rule, in such a way that  these two terms have been used interchangeably. Another mechanism of generating highly-skewed statistical  distributions is through  multiplicative random walk or autocatalytic process. This mechanism is frequently  associated with the log-normal degree distribution.\cite{Mitzenmacher2004,Caldarelli2007}

Usually, the conclusion on whether a  complex network is generated by this or that mechanism is drawn as follows. A researcher assumes a microscopic mechanism of network growth  and measures degree distribution of this network. In  most cases he finds some power-law  distribution, measures its  exponent, compares it to model prediction, and decides whether his observations validate the suggested mechanism or not. This procedure hinges on the conclusion whether the degree distribution of this network is a power-law or something else. To draw such  conclusion is not easy. Clauset, Shalizi, and Newman \cite{Clauset2009} analyzed many  complex networks and suggested a series of stringent mathematical tests to discriminate between the power-law and other heavy-tail degree distributions. While many of the analyzed networks were previously claimed to have the power-law degree distribution, Ref. \cite{Clauset2009} found  them  in less than a half of these networks. This sobering assessment resulted in a wave of criticism questioning the ubiquity of the power-law degree distributions in complex networks and their scale-free character. In particular, Ref. \cite{Stumpf2012} insists that most  reported power-law distributions lack statistical support and mechanistic backing, Ref. \cite{Lima2009} claims that the power-law  and the scale-free distribution are not the same, Ref. \cite{Willinger2009} dismisses the myth of scale-free networks. Thus, the ubiquity of power-law distributions in complex networks and  their significance for pinpointing the mechanism of network growth is now under question.  To resolve this question  Mitzenmacher \cite{Mitzenmacher2005}  suggested to leave attempts of derivation of the network generating mechanism  from degree distributions and  to measure this mechanism directly, for example, by applying the time-series analysis.

In our recent study \cite{Golosovsky2017} we closely followed this suggestion. Namely, we took a well-defined complex network - citation network of Physics papers- and established  its microscopic growth mechanism using two complementary time-resolved techniques: (*) analysis of the age composition of the reference lists of  papers (retrospective approach), and (**) analysis of citation trajectories of individual papers (prospective approach). As expected, both approaches yielded consistent results and allowed us to formulate a stochastic model of citation dynamics of individual papers. The parameters of the model were found from the measurements of citation trajectories of individual papers rather than from citation distributions. Quite unexpectedly, our measurements revealed that citation dynamics is nonlinear.

In our present study  we address the following question: what is the functional form of  degree distribution in citation networks? We approach this question from two directions. First, we chose several well-defined citation networks and measured how their degree distributions evolve with time. We analyzed these distributions empirically using commonly accepted strategies. Surprisingly, we found that citation distributions are non-stationary and do not converge to some limiting distribution, even in the long time limit. This nonstationarity explains  why previous efforts to derive  growth mechanism of citation networks from  degree distribution were so inconclusive. Second, we modeled these citation distributions using our nonlinear stochastic model of citation dynamics \cite{Golosovsky2017} and found explanation for the nonstationarity.  It turns out that  nonstationarity of  citation distributions and their "power-law" shape both originate in  \emph{nonlinear} citation dynamics. The nonlinearity introduces a certain scale, in such a way that  citation networks can no longer be considered scale-free.

\section{Empirical analysis of citation distributions}
\label{sec:measurements}
\subsection{Measurement of citation distributions}

\begin{figure}[!ht]
\begin{center}
\includegraphics*[width=0.45\textwidth]{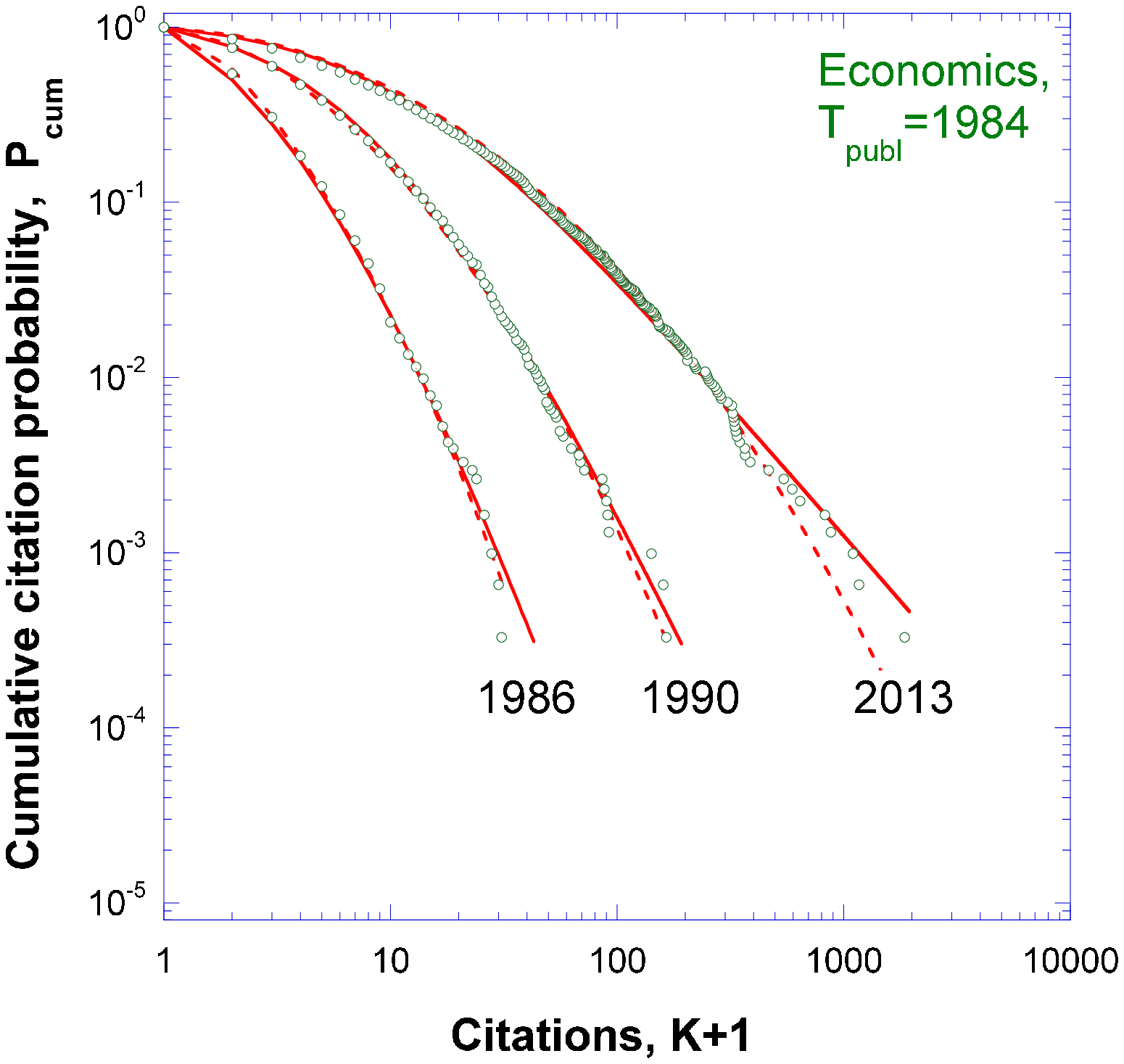}
\includegraphics*[width=0.45\textwidth]{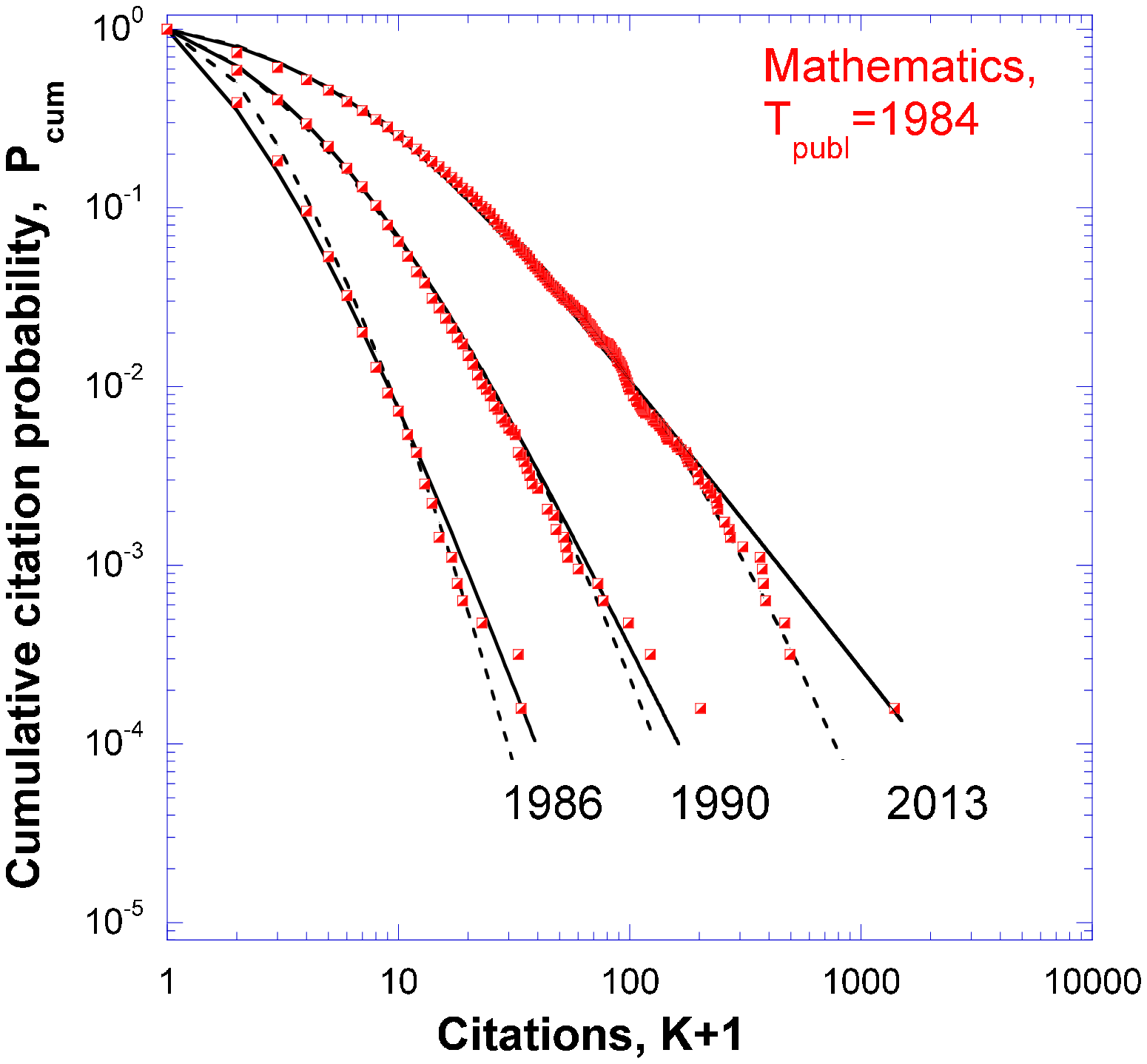}
\includegraphics*[width=0.45\textwidth]{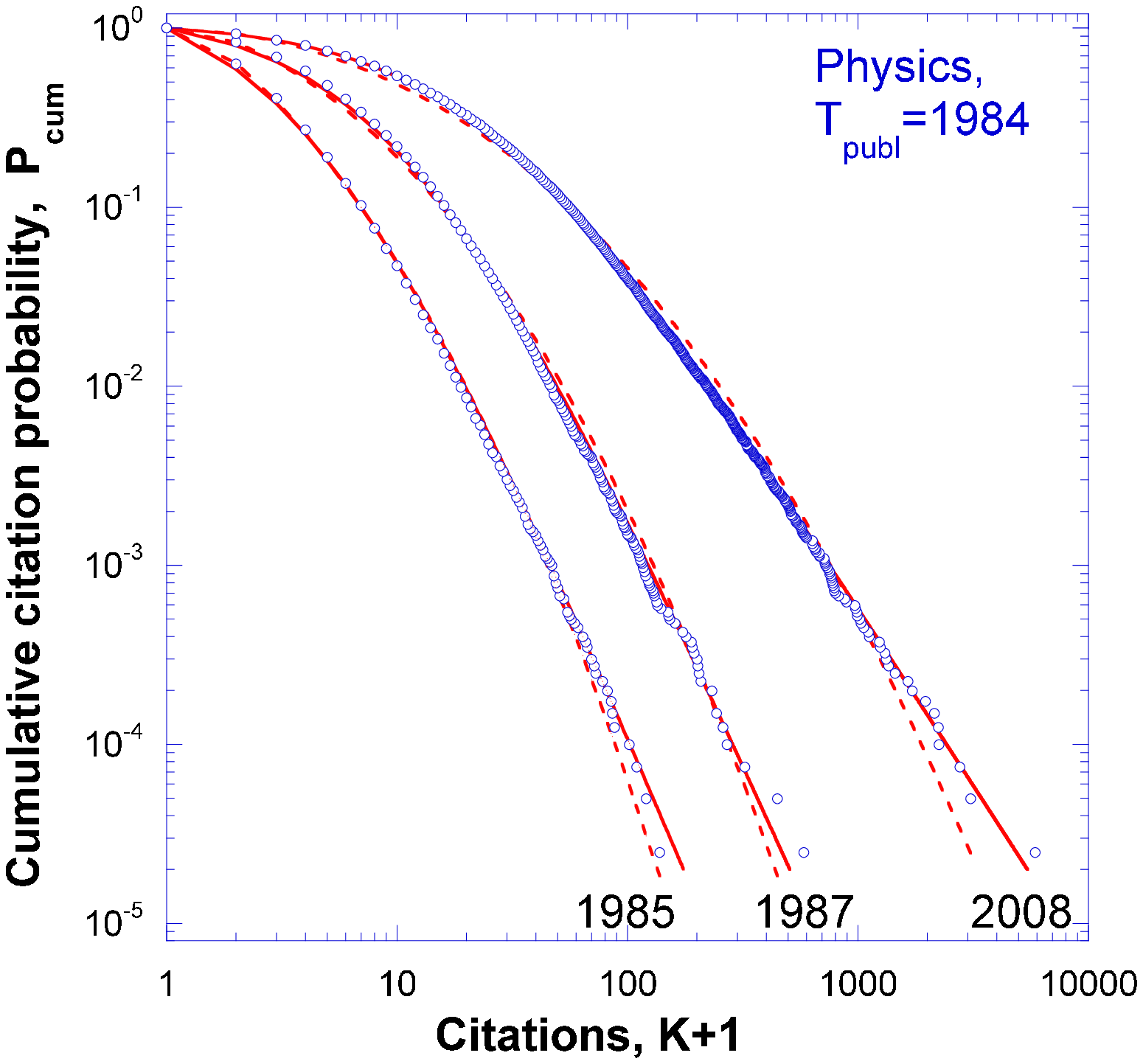}
\caption{Cumulative citation distributions  for the original research papers published in 1984. The points stay for measurements, the continuous lines show the discrete power-law  fit (Eq. \ref{Waring}), the dashed lines show the log-normal  fit (Eq. \ref{lognormal}). (a) 3043 Economics papers, (b) 6313 Mathematics papers, (c) 40195 Physics papers.
}
\label{fig:cdf}
\end{center}
\end{figure}

To find the functional form of citation distributions we chose several well-defined citation networks and focused on all original research papers (overviews excluded) published in one year  in one field.\cite{community}  In particular, we considered the fields of Physics, Mathematics, and Economics and the publication year 1984.\cite{year1984} We used the Thomson Reuters ISI Web of Science   and  found 40195 Physics papers, 6313 Mathematics papers, and 3043 Economics papers published in 1984. We measured $K(t)$, the cumulative number of citations garnered by each of these  papers during subsequent $t$ years, the publication year corresponding to $t=1$. For each field, the fraction of  papers  that garnered $K$ citations by the year $T_{publ}+t-1$ yields the  probability density function, $p(K,t)$. Figure \ref{fig:cdf} shows the corresponding cumulative citation distributions,  $P(K,t)=\sum_{j=K}^{\infty}p(j,t)$. These distributions are highly-skewed.

We fitted these distributions  using discrete power-law (Eq. \ref{Waring}) and discretized  log-normal function (Eq. \ref{lognormal}). The former fit assumes a straight tail in the log-log plot while the latter fit assumes a convex tail. Figure \ref{fig:cdf} shows that for small $t$ (early after publication) both fits perform equally well, while for later years the discrete power-law fit is better. Indeed, for the most representative  set of Physics papers the tail of the distribution is straight, as suggested by Eq. \ref{Waring}, rather than convex, as suggested by Eq. \ref{lognormal}. In what follows we use the discrete power-law (Eq. \ref{Waring}) to parameterize  these citation distributions.

\begin{figure}[!ht]
\begin{center}
\includegraphics*[width=0.47\textwidth]{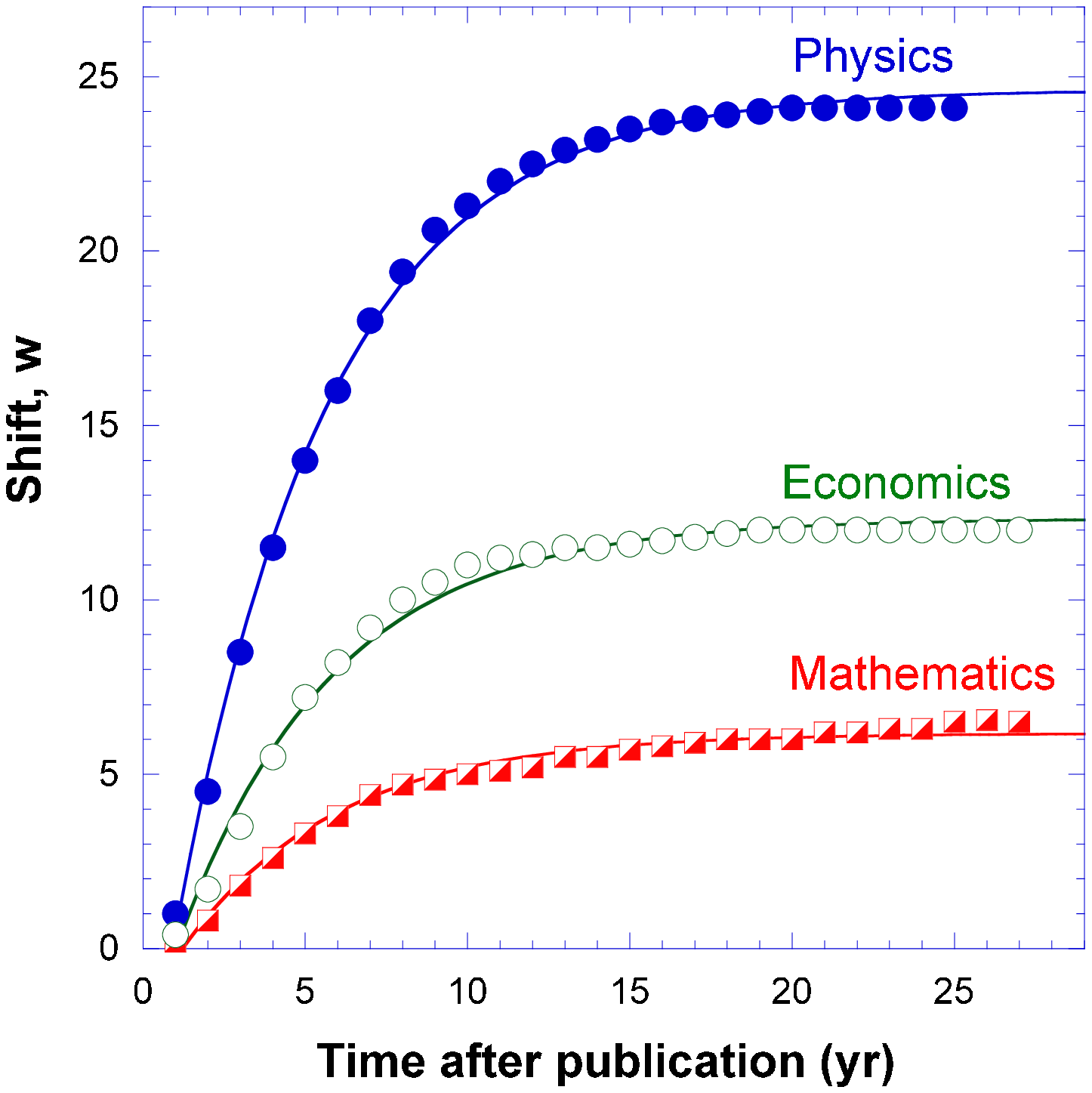}
\includegraphics*[width=0.47\textwidth]{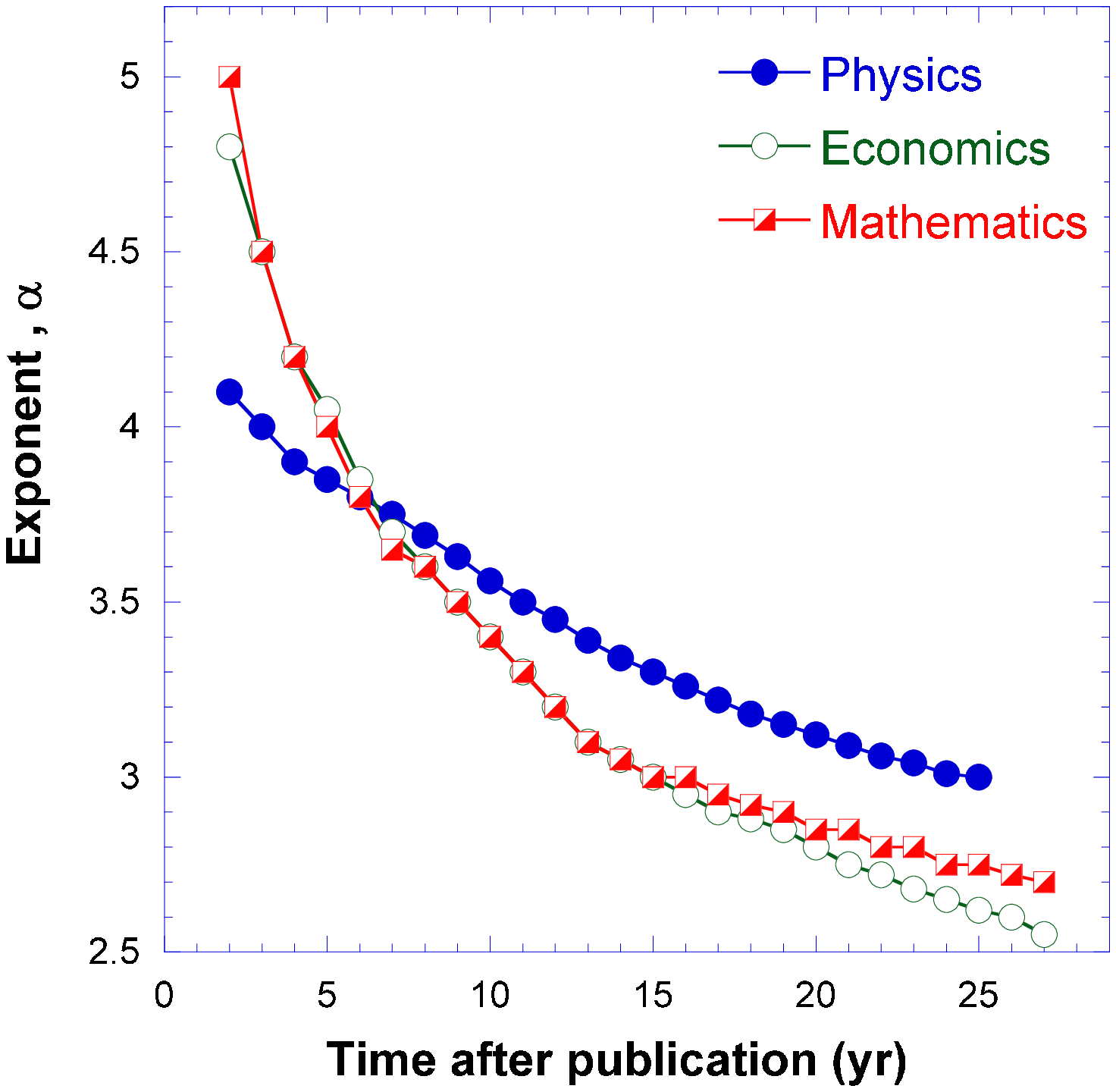}
\caption{Parameters of the  discrete power-law fit (Eq. \ref{Waring}). (a) Shift. (b) Exponent. Although  the shift $w$ comes to saturation after 10 years, the exponent  $\alpha$ continuously decreases  and does not come to saturation even after 25 years.
}
\label{fig:Waring}
\end{center}
\end{figure}

Figure \ref{fig:cdf} shows that, as time passes,  citation distributions shift to the right and the slope of their tails  becomes more gradual.   Figure \ref{fig:Waring}  shows time dependence of the fitting parameters $w$ and  $\alpha$  that capture, correspondingly, the shift and the slope.  The shift $w$ increases with time and, for all three fields, it comes to saturation after $\sim$10 years. The exponent $\alpha$ continuously decreases but  does not come to saturation. This means that  even after 25 years  citation distribution is not stationary and its tail still develops.

\subsection{Mean number of citations}
\begin{figure}[!ht]
\begin{center}
\includegraphics*[width=0.5\textwidth]{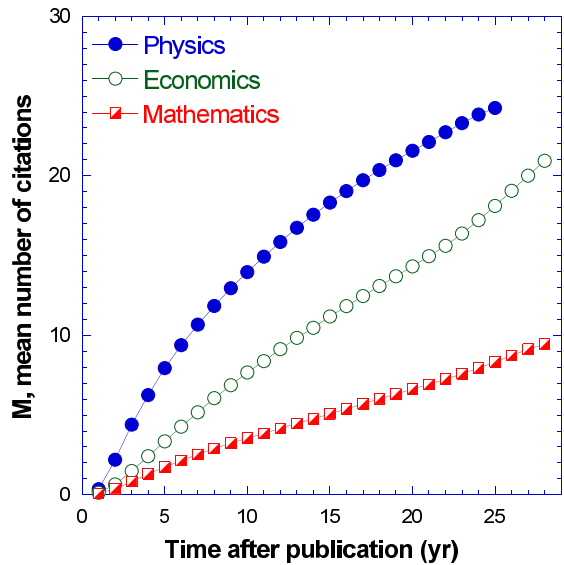}
\caption{$M$, the mean number of citations, does  not come to saturation even after 25 years.
}
\label{fig:mean}
\end{center}
\end{figure}

Another indication of nonstationary of  citation distributions comes from the analysis of the mean number of annual citations, $M(t)=\sum_{K=1}^{\infty}Kp(K,t)$.  If citation distributions were converging to some limiting shape, then the mean of the distribution should saturate in the long time limit. Figure \ref{fig:mean} shows that $M(t)$ does not come to saturation for either field. Moreover, for Mathematics and Economics papers  $M(t)$ grows with acceleration!

\begin{figure}[!ht]
\begin{center}
\includegraphics*[width=0.6\textwidth]{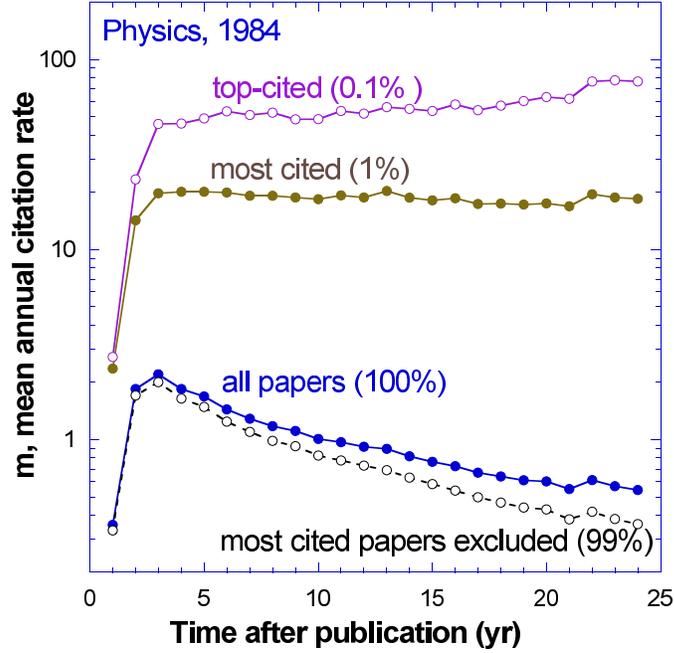}
\caption{Mean annual citation rate of three sets of Physics papers  published in 1984: 40 top-cited papers (violet circles), 400 most cited papers (brown circles), all 40195 Physics  (blue circles). The open black circles stay for the mean citation rate of all Physics papers excluding most cited ones.  While  citation rate of an ordinary paper decays almost to zero after 20 years, citation rate of highly-cited papers does not decay and  even grows with time.
}
\label{fig:cit-mean-dynamics}
\end{center}
\end{figure}

To understand  why $M(t)$  does not come to saturation we considered its constituents  in more detail. To this end we focused on Physics papers which make a biggest set. We ranked these papers according to the number of citations garnered after 25 years. Then we arranged them into three overlapping sets:  40 top-cited papers, 400 most cited papers, and all 40195 Physics papers. For each set we  measured $m=dM/dt$, the mean annual number of citations (citation rate). Figure \ref{fig:cit-mean-dynamics} shows the  $m(t)$ dependencies. They represent an average citation trajectory of a top-cited, highly-cited, and ordinary paper, correspondingly. These trajectories are qualitatively different. While the mean citation rate of an ordinary paper  grows fast during 1-2 years after publication and then slowly decays (obsolescence), the mean citation rate of the highly-cited papers  does not decay, and that of the top-cited papers even accelerates with time.  In other words, while citation career of an ordinary paper eventually comes to saturation, the highly-cited papers are cited permanently, they are immortal.

We can also ask a following question: what is the contribution of the highly-cited papers to the mean citation rate? By analyzing citation dynamics of the papers published in five leading scientific  journals in 1990  Barabasi, Song, and Wang \cite{Barabasi2012}  found that $1\%$ of top-cited papers garnered disproportionately high fraction of citations after 20 years.  To see whether this conclusion holds  for  Physics as well, we compared the mean citation rate  for  two following sets of papers: (*) all 40195 Physics papers published in 1984  (100$\%$, blue circles), and (*) all Physics papers excluding $1\%$ of most cited papers (99$\%$, open black circles). The difference between  these two sets represents contribution of $1\%$ most cited papers. While for the first 5 years after publication the fraction of citations garnered by most cited papers is small and stays in proportion to their low number, for later years this proportion is disproportionately high.  In particular, in the 25-th year after publication  44$\%$  of all annual citations in Physics  come from 1$\%$ of the papers.

\subsection{Citation lifetime}
\begin{figure}[!ht]
\begin{center}
\includegraphics*[width=0.5\textwidth]{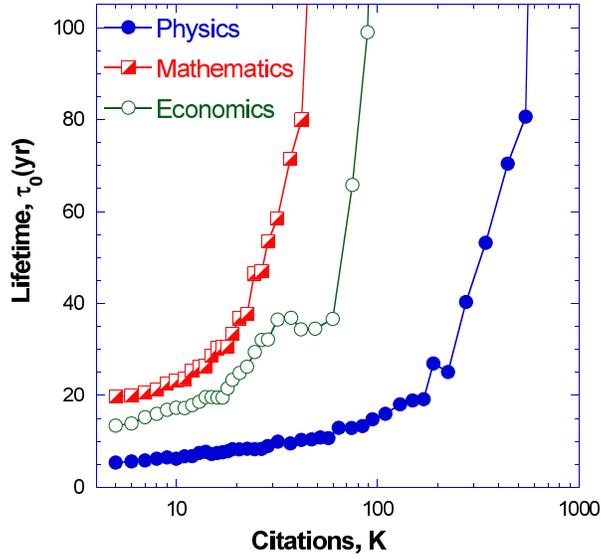}
\caption{Citation lifetime $\tau_{0}$ versus  $K$, the number of citations after 25 years (the long time limit of citations). The measurements were taken for the papers shown in Fig. \ref{fig:cdf}. $\tau_{0}$ grows with increasing $K$ and diverges at some $K_{0}$. The highly-cited papers with $K>K_{0}$ are runaways.  The solid lines are the guide to the eye.
}
\label{fig:lifetime}
\end{center}
\end{figure}

The dichotomy between the low-cited papers with their decaying  citation rate and the highly-cited papers with their increasing citation rate, as it is demonstrated in Fig. \ref{fig:cit-mean-dynamics}, is not rigid, there is a continuous transition between these two classes of papers. To show this we  analyze the paper's longevity in a way similar to that used in our earlier publication.\cite{Golosovsky2013} We approximated citation trajectory of each paper  by an exponential dependence  $K=K_{\infty}(1-e^{-\frac{t-\Delta}{\tau_{0}}})$ where $\tau_{0}$ is the citation lifetime, $K_{\infty}$ is the number of citations in the long time limit, and $\Delta\sim 1-2$ years characterizes a delay in citation career of the paper. Figure \ref{fig:lifetime} plots $\tau_{0}$  versus $K$, the number of citations after 25 years, which we take as a substitute for $K_{\infty}$.\cite{lifetime}   We observe that $\tau_{0}$ grows continuously with $K$ and diverges at certain $K_{0}$, in such a way that the papers with $K>K_{0}$  exhibit runaway behavior - their citation career does not saturate.

\begin{figure}[!ht]
\begin{center}
\includegraphics*[width=0.6\textwidth]{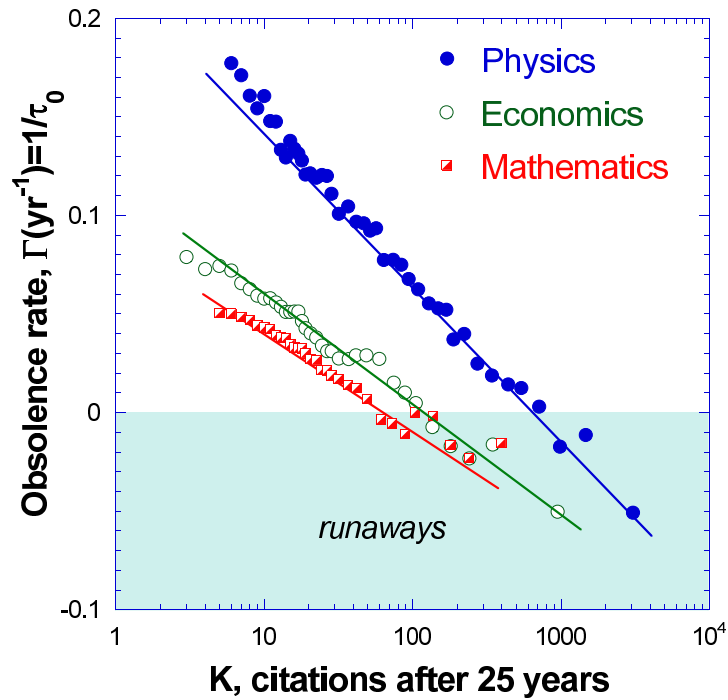}
\caption{The obsolescence rate, $\Gamma=\tau_{0}^{-1}$ versus long time limit of citations for the papers published in 1984. For each field  $\Gamma$ decreases (lifetime increases) with the number of citations. Above certain $K_{0}$ (760 for Physics, 113 for Economics, and 55 for Mathematics) the obsolescence rate  $\Gamma$ changes sign indicating the onset of the runaway behavior. The solid lines show empirical logarithmical dependence   given by Eq. \ref{logarithm}.
}
\label{fig:obsolescence}
\end{center}
\end{figure}

To include these runaways in our discussion we considered the obsolescence rate $\Gamma=1/\tau_{0}$. Figure \ref{fig:obsolescence}  shows that  $\Gamma$ decreases logarithmically  with $K$,
\begin{equation}
\Gamma=\Gamma_{0}-b\ln{K},
\label{logarithm}
\end{equation}
where  $\Gamma_{0}$ and $b$ are parameters which depend on the field and publication year. The function $\Gamma(K)$ changes sign and becomes negative at certain $K_{0}$. Negative obsolescence rate indicates exponentially increasing number of citations- the runaway behavior. Thus, the papers with $K<K_{0}$ have finite lifetime and eventually become obsolete, the papers with $K>K_{0}$ are immortal - their citation career can continue indefinitely.   $K_{0}$ is found from the following relation: $\ln{K_{0}}=\Gamma_{0}/b$ and it sets scale for citation distributions. We found that $K_{0} =760$ citations for Physics,  113 citations for Economics, and 55 citations for Mathematics papers published in 1984.

The parameter $\Gamma_{0}$  defines  longevity of the ordinary papers. Indeed, for small $K\sim 1$, Eq. \ref{logarithm} reduces to $\tau_{0}=\Gamma_{0}^{-1}$. This yields $\tau_{0}=4.6,9$, and 11.8 years for Physics, Economics, and Mathematics, correspondingly. Since citation trajectory of the ordinary papers is close to exponential, it comes to saturation after $3\tau_{0}$. The longer citation lifetime of the Economics and Mathematics papers, as compared to that of Physics papers, is related to propensity of these fields to cite old papers and to more rapid growth of  the number of papers as covered by Web of Science database.

\subsection{Summary of measurements}
Our empirical observations can be summarized as follows. The early citation distributions (Fig. \ref{fig:cdf}) can be fitted either by the log-normal or by the discrete  power-law  distribution with $\alpha=4-5$ (Eq. \ref{Waring}). Large exponent indicates that these distributions  are  "conventional" and their tails play only a minor role in defining the mean and the variance of the distribution. As time passes and  papers garner more citations, citation distributions  shift to the right. This shift mostly comes to an end after $\sim$ 10 years when citation career of ordinary papers is over. Later on, citations are garnered mostly by the highly-cited papers which compose the tail of the distribution. The slope of the distribution becomes more gradual  since, as time passes, the tail moves fast to the right while the rest of the distribution is slowed down.  This is the reason why the power-law exponent $\alpha$ decreases with time (Fig. \ref{fig:Waring}). Eventually, the tail of the distribution comprises only runaway papers whose citation career continues indefinitely. While the tail  continues to move to the right, the rest of the distribution stays immobile, in such a way that citation distribution never comes to saturation.

Thus, while citation distributions can be fitted by the discrete power-law, they are nonstationary. Although at each temporal snapshot citation distribution can look as a scale-free, there is a certain dynamic scale $K_{0}$ which can be inferred from citation trajectories.  In what follows we dwell into microscopic mechanism responsible for the temporal evolution of citation distributions.

\section{Nonlinear stochastic model of citation dynamics explains all observed features of citation distributions}
\subsection{Model}
\begin{figure}[!ht]
\includegraphics*[width=0.6\textwidth]{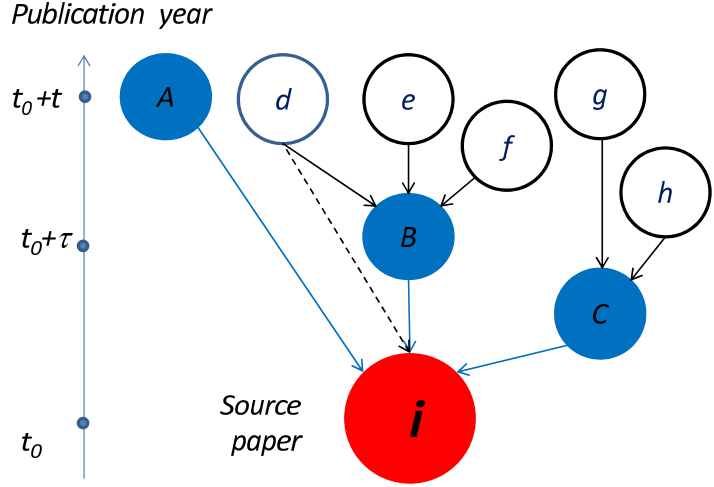}
\caption{A fragment of citation network showing a source paper \emph{i} and its citing papers.  The papers \emph{A,B,C} are direct citing papers since they cite \emph{i} and do not cite any other paper citing \emph{i}.  The paper \emph{d} cites both \emph{B} (which cites \emph{i})  and  \emph{i}, therefore, it is indirect citing paper.  The solid and dashed lines link the source paper  with its  direct and indirect citing papers, correspondingly. Each indirect citing paper closes a triangle  in which the source paper \emph{i} is one of the vertices. The papers \emph{A,B,C,d} cite the source paper  \emph{i} and they are the first-generation citing papers.  The  papers \emph{d,e,f} and \emph{g,h}  are second-generation citing papers since they cite  the first-generation citing papers \emph{B} and \emph{C}, correspondingly.  The indirect citing papers, such as \emph{d}, belong to both generations. The vertical scale shows publication year of each paper.
}
\label{fig:model}
\end{figure}

We have recently developed a stochastic model of citation dynamics of scientific papers.\cite{Golosovsky2017}  The model is based on the triadic closure/copying/redirection mechanism which is schematically shown in Fig. \ref{fig:model}. In what follows we briefly recapitulate our model. It assumes that each published paper (we name it source paper) generates two kinds of citations:  direct citing papers whose authors find it in databases or Internet, and indirect citing papers whose authors learn about it from the reference lists of already selected papers (and copy it to their reference list).
The model assumes that  the probability of a paper $i$ to garner $k$ citations in year $t$ after its publication follows a Poisson distribution, $Poiss(k)=\frac{(\lambda_{i})^{k}}{(k)!}e^{-\lambda_{i}}$, where $\lambda_{i}$ is the paper-specific latent citation rate. This rate is given by the following expression:
\begin{equation}
\lambda_{i}(t)= \lambda_{i}^{dir}(t)+\sum_{\tau=1}^{t}P_{i}(t-\tau)N(t-\tau)k_{i}(\tau).
\label{model}
\end{equation}
where $\lambda_{i}^{dir}(t)$ is the direct citation rate and $t$ is the time after publication of the source paper. The second addend stays for indirect citation rate.

The first addend, $\lambda_{i}^{dir}(t)$, captures dynamics of direct citations.  The total number of those that the paper \emph{i} garners in the long time limit is $\eta_{i}=\int_{0}^{\infty}\lambda^{dir}_{i}(t)dt$  where $\eta_{i}$ is called paper's fitness. Our definition of fitness is different from that of Bianconi and Barabasi \cite{Bianconi2001} and is more close to that of Caldarelli et al. \cite{Caldarelli2002}. By measuring citation trajectories of individual Physics papers and keeping distinction between the direct and indirect citations, we found that $\lambda_{i}^{dir} =\eta_{i}m_{dir}(t)$,  where $m_{dir}(t)$ is an empirical function,  the same for all papers in one field published in one year.\cite{Golosovsky2017} This function is shown in SM and, given our definition of $\eta_{i}$,  it satisfies  condition $\sum_{t=1}^{\infty}m_{dir}(t)=1$.

The second addend in Eq. \ref{model} stays for indirect citations. Here, $k_{i}(\tau)$ is the total number of citations that the paper $i$ garnered in year $\tau$ after publication. $k_{i}(\tau)$ is equal to the number of the first-generation citing papers published in year $\tau$. Each of them generates a train  of second-generation citing papers published later at $t>\tau$. We denote by  $N(t-\tau)$  the average  number of the latter published in year $t$ that were generated by one first-generation citing paper published in year $\tau$. Each of these $N(t-\tau)$ second-generation citing papers can cite (indirectly) the source paper $i$ with probability  $P_{i}(t-\tau)$.

We found functions $N(t-\tau)$ and $P_{i}(t-\tau)$ by measuring citation trajectories of individual Physics papers and by counting their first- and second-generation citations.\cite{Golosovsky2017} These studies showed that the  function $N(t-\tau)$ is almost the same for all source papers published in one year while the probability of indirect citation (copying) is paper-specific and is captured by the empirical expression $P_{i}(t-\tau)=P_{0}e^{-\gamma(t-\tau)}$ where $\gamma$=1.2 yr$^{-1}$. Quite unexpectedly, we found that $P_{0}$ is not constant but depends on the number of previous citations of the source paper, this dependence could be traced to the assortativity of citation networks. For Physics papers published in 1984 we found that
\begin{equation}
P_{0}(K_{i})=0.25(1+0.36\ln K_{i}).
\label{nonlinearity}
\end{equation}
This $P_{0}(K)$ dependence  introduces nonlinearity in Eq. \ref{model}. In what follows we demonstrate that this nonlinearity is the source of all interesting features of citation distributions.

\subsection{Citation distributions are nonstationary}
We used Eq. \ref{model} to simulate citation trajectories of the Physics papers published in 1984. With the exception of $\eta_{i}$ all other parameters  in this equation are not paper-specific and were measured from citation trajectories of papers and not from citation distributions.  To run numerical simulation we need initial condition for each paper and this condition is set by $\eta_{i}$, paper's fitness. To assign a certain fitness to each paper we used the following consideration.  Figure \ref{fig:model} shows that indirect citations lag in time after direct citations. Since the characteristic time for publishing a paper is one year, the minimal time lag between the publication of the first indirect citing paper and its source paper is around two years. Therefore, during first couple of years after publication of the source paper, citations are mostly direct. Hence, by measuring the number of citations a paper garnered during first couple of years after publication we can estimate its fitness from the relation $K_{i}(t=2)\approx\eta_{i}[m_{dir}(t=1)+m_{dir}(t=2)]$.

\begin{figure}[]
\includegraphics*[width=0.47\textwidth]{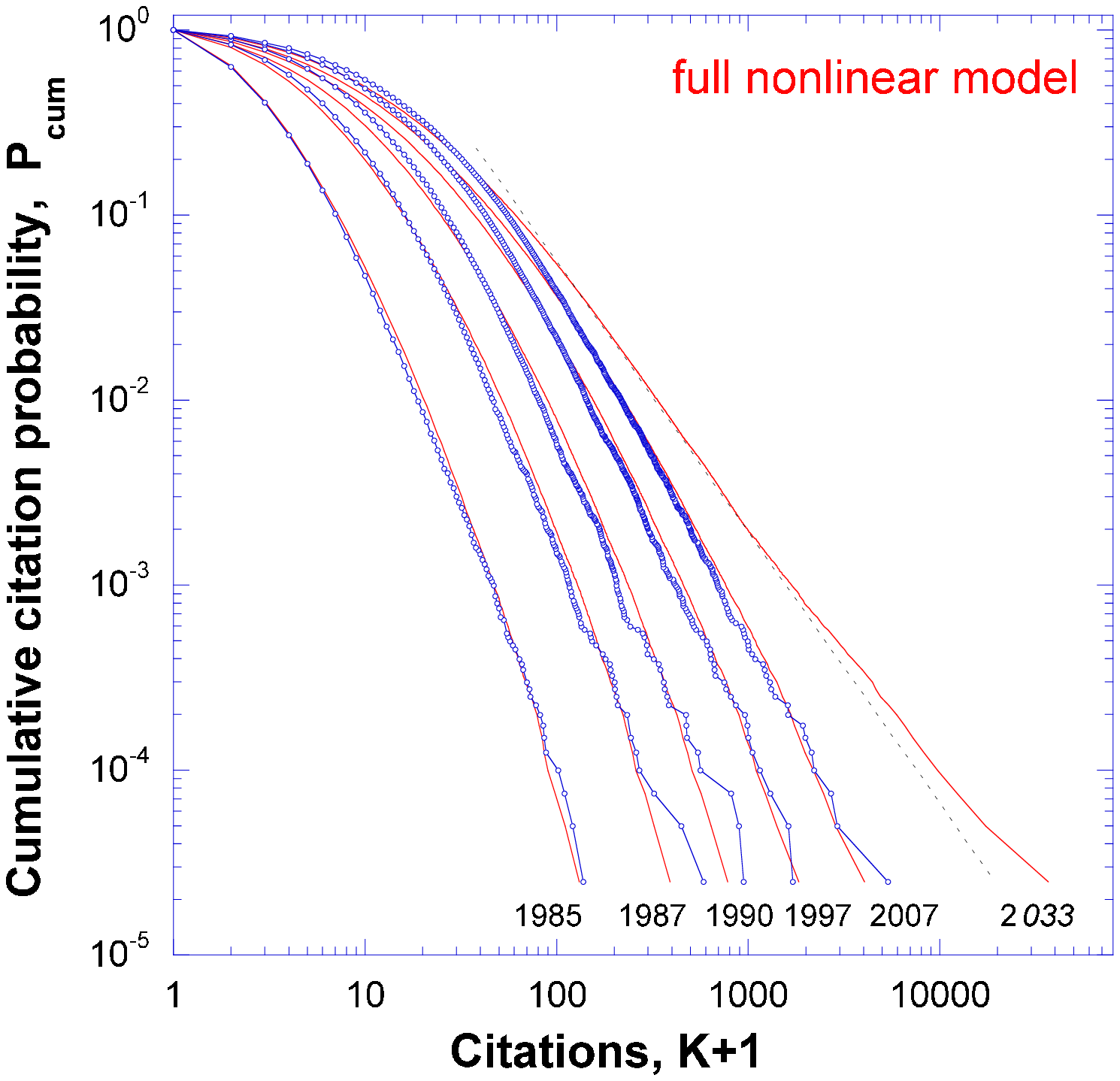}
\includegraphics*[width=0.47\textwidth]{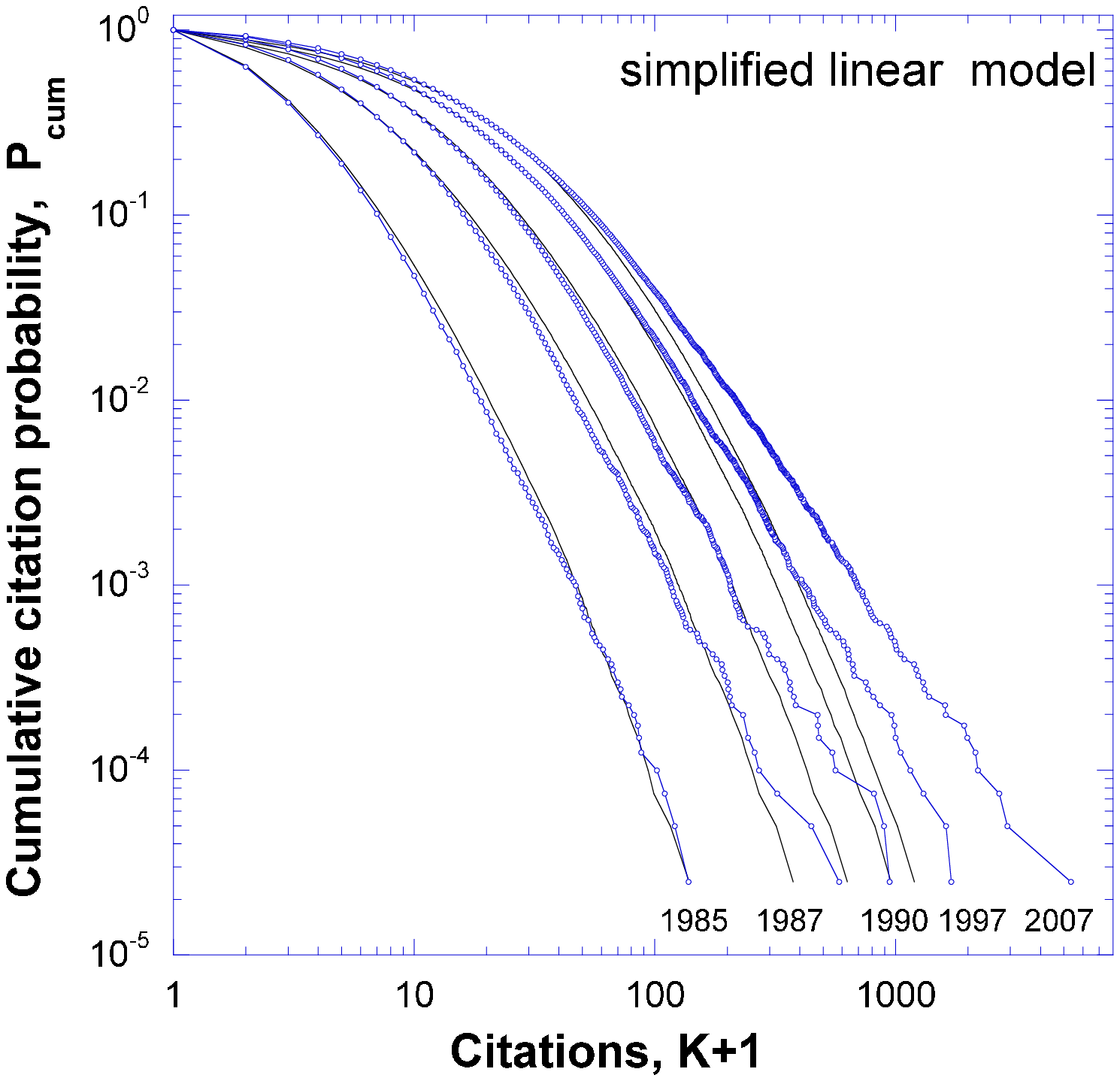}
\caption{Cumulative citation distributions for 40195 Physics papers published in 1984.  The circles stay for our measurements, continuous lines stay for stochastic numerical simulation  based on Eq. \ref{model} (the average over 30 realizations).  (a)  Simulation based on Eq. \ref{model} with the  kernel given by Eq. \ref{nonlinearity} (full nonlinear model). Note excellent agreement between the simulation and measurements.  Continuous red line, that shows extrapolation to the year 2033, is remarkable. While the tails of the distributions for 1984-2007 look as straight lines, the simulated  2033 distribution exhibits concave tail.  This is clearly seen by comparing it to a dashed line which is a linear extrapolation of the straight part of the distribution for $100<K<1000$. The concave tail is an indicator of the runaway behaviour. (b) Simulation based on  Eq. \ref{model} with a constant kernel, $P_{0}=0.54$ (a  linear model). The tails of the simulated distributions  are almost straight lines with  time-independent slope. While the linear model fits the small-degree part of the distributions (low-cited papers), it fails to match the tails (highly-cited papers).
}
\label{fig:long-simulation}
\end{figure}

An almost equivalent approach  consists in using $\eta_{i}$ as a fitting parameter for each paper and running the numerical simulation with the aim of fitting citation distributions for the first 2-3 years. The result of this fitting procedure is the fitness distribution. Then, with this fitness distribution, we can run the model for later times without additional fitting parameters.   Figure \ref{fig:long-simulation}a shows an excellent agreement between the measured and numerically simulated citation distributions where fitness distribution  is a log-normal  with $\mu=1.62$ and $\sigma=1.1$.\cite{Golosovsky2017}   Thus, the agreement between the numerical simulation and the early citations distributions is built in, but the agreement  with the late citation distributions  is nontrivial. Given this agreement we extrapolated our simulation to the future, up to year 2033. Figure \ref{fig:long-simulation}a shows that numerically simulated citation distributions do not become stationary and continue to develop  even 50 years after publication. This explains our empirical observation of nonstationary citation distributions  as evidenced by Figs. \ref{fig:Waring}, \ref{fig:mean}.

Which feature of our model is responsible for nonstationary citation distributions?  In what follows we prove that this is the $P_{0}(K)$  dependence (Eq. \ref{nonlinearity}) which drives dynamics of indirect citations.  To demonstrate this we performed numerical simulation where instead of Eq. \ref{nonlinearity} we set $P_{0}=0.54$. This renders our model  linear.  Figure \ref{fig:long-simulation}b shows that the linear model accounts for citation dynamics of the papers with $K<100$  for all times, while citation dynamics of the papers with $K>100$ is captured only at early times. This is because the linear model captures dynamics of direct citations and fails to account for dynamics of indirect citations which play a major role for highly-cited papers. From another perspective, Fig. \ref{fig:long-simulation}b shows that while the linear model accounts fairly well for  early citation distributions, it fails miserably for  late distributions. Simulated citation distributions shift with time to the right while their slopes do not change.  This does not match our measurements which indicate that, as time passes,  the slope of citation distributions continuously decreases. We conclude that the nonstationarity and decreasing slope of citation distributions  are consequences of the $P_{0}(K)$  dependence given by Eq. \ref{nonlinearity}.


\subsection{Citation distributions do not necessarily follow the power-law dependence}
Since the measured citation distributions have been successfully fitted by the discrete power-law (Fig. \ref{fig:cdf}, Eq. \ref{Waring}) we turn to our model for the justification of such fit. To this end we studied which parameters of the model are responsible for the general shape of citation distributions. We notice that early citation distributions mimic the fitness distribution. For the Physics papers these can be represented either as  log-normal with $\mu=1.62,\sigma=1.1$ or Waring distribution with $\alpha=4$. Anyway,  both these distributions are convex, especially at low $K$. Numerical simulation shows that, as time passes, citation distribution shifts to the right and its tail straightens, in such a way that it starts to look  as a power-law dependence with $2<\alpha<3$. This power-law-like distribution holds for 3-20 years after publication and this is the reason of why citation distributions are successfully fitted by the discrete power-law (Eq. \ref{Waring}).  However, for longer time windows and for large sets of papers the situation starts to change. Extrapolation of the simulation to $t=50$ years shows that, in the long time limit, the distribution becomes concave (Fig. \ref{fig:long-simulation}a, the line corresponding to year 2033) rather than remaining straight.  This means that the straight  tail of citation distributions at intermediate times is probably only a transient shape.

Nevertheless, the overwhelming majority of the reported  citation distributions have straight or slightly convex tails. So,  why concave tail is so rare?  Our results suggest that to observe a concave tail of the citation distribution there shall be long time window and large dataset that contains many runaway papers.   Our set of all 40195 Physics papers published in one year  is still insufficient for this purpose. However, the 10-times bigger set of 418,438 Physics papers published in 1980-1989 does reveal the concave tail in the citation distribution, as we have showed  in our previous publication.\cite{Golosovsky2012a} In SM we show citation distribution for this extremely big set of papers along with a simple analysis.



\subsection{Citation distributions are not scale-free}
Previous empirical studies, which claimed the power-law citation distributions,  implied that these distributions are scale-free.\cite{Dorogovtsev2001,Clauset2009,Caldarelli2007} Of course, since citations are discrete and non-negative, citation distributions have a natural scale- the mean number of citations $M$. By a  "scale-free citation distribution" one usually means the absence of the macroscopic scale besides the microscopic scale set by $M$. In particular, for Physics papers published in 1984, the mean number of citations in 2008 is $M=26$.  This scale is visible in Fig. \ref{fig:cdf}c and it corresponds to transition from the curved part of the distribution  to the straight tail. This straight tail extends from $K/M\sim3$ to  $K/M=230$. The huge disparity between these two numbers  is usually considered as an indicator of the scale-free behavior. However, this is only an indicator and  not a proof. In what follows we demonstrate that citation distributions do have a scale. This hidden scale is barely visible in citation distributions but it pops out explicitly when we analyze citation trajectories of the papers. We have demonstrated this scale $K_{0}$ in our measurements, now we show it in the simulations.

\begin{figure}[!ht]
\begin{center}
\includegraphics*[width=0.55\textwidth]{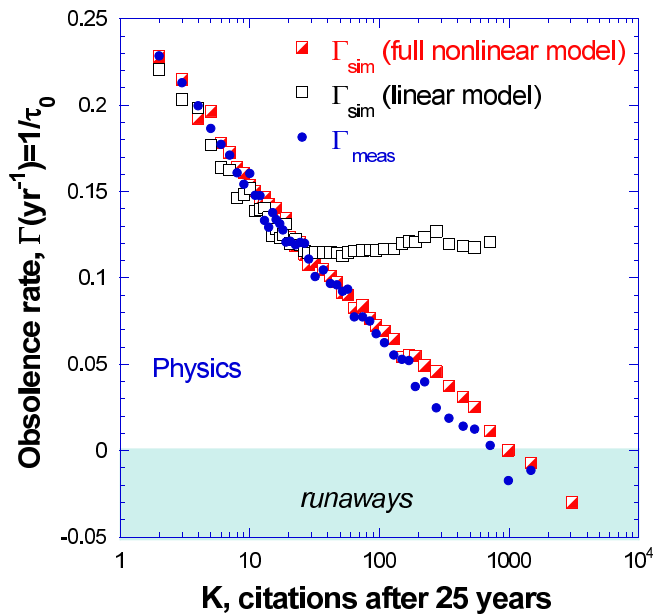}
\caption{The obsolescence rate of Physics papers $\Gamma=1/\tau_{0}$ versus $K$, the number of citations after 25 years.  Blue circles show the results of our measurements from Fig. \ref{fig:obsolescence}. Red squares show results of simulations  based on  full nonlinear model with the  kernel given by Eq. \ref{nonlinearity}.  $\Gamma$ decreases with increasing $K$ and changes sign  at $K_{0}$.  This decreasing trend is in agreement with our measurements.  Open black squares show results of simulations  based on  a linear model with a constant kernel, $P_{0}=0.54$. $\Gamma$ decreases with $K$ and achieves a constant level at $K\approx 20$.  This simulation disagrees with our measurements for $K>20$.
}
\label{fig:lifetime1}
\end{center}
\end{figure}

To this end we come back to our numerical simulation for 40195 Physics papers published in 1984  and  focus on citation trajectories of individual papers. By analyzing  these trajectories we found citation lifetime $\tau_{0}$ in the same way as we did in Section \ref{sec:measurements} for the measured citation trajectories. Figure \ref{fig:lifetime1} shows the corresponding obsolescence rate $\Gamma_{sim}= 1/\tau_{0}$. The simulated $\Gamma_{sim}$  agrees  fairly well with  $\Gamma_{meas}$. Thus our model reproduces fairly well the decreasing $\Gamma(K)$ dependence and the divergence of citation lifetime at certain $K_{0}$.


Which feature of the model is responsible for this surprising $\Gamma(K)$ dependence? We claim that this is again the  $P_{0}(K)$ dependence (Eq. \ref{nonlinearity}). Indeed, numerical simulation using a linear model with $P_{0}=const$ (Fig. \ref{fig:lifetime1}, open circles) shows that $\Gamma$ decreases with $K$ only at small  $K<20$.   This $\Gamma(K)$ dependence  arises from the fact that citations are discrete. Indeed,  Eq. \ref{model} describes a discrete Hawkes process where citation rate of a paper depends mostly on the recent citation rate. This dependence introduces a positive feedback that amplifies fluctuations. For $K>20$ this purely statistical effect is washed out. While the linear model accounts for our measurements only for $K<20$ and disagrees with them for  $K>20$, the nonlinear model accounts for our measurements fairly well for all $K$.

We conclude that decreasing $\Gamma(K)$ dependence for $K>20$ is a direct evidence for the $P_{0}(K)$ dependence. Since the measured $\Gamma(K)$ dependences for the Mathematics, Economics papers are very similar to that for Physics papers (Fig. \ref{fig:obsolescence}), we infer  that citation dynamics of the Economics and Mathematics papers (Fig. \ref{fig:cdf}) also follow nonlinear model with $P_{0}(K)$ dependence described by Eq. \ref{nonlinearity}, albeit with different coefficients.

\section{Continuous approximation of the model}
To  better understand  how $\Gamma(K)$ results from the $P_{0}(K)$ dependence we analyze  continuous approximation of our model.   Namely, we  disregard stochasticity and replace the latent citation rate $\lambda_{i}$ in Eq. \ref{model} by the actual citation rate $k_{i}$ which is considered as a continuous variable. The time is also continuous, hence we replace the sum by the integral.  Due to strong exponential dependence of $P_{i}(t-\tau)$ and much weaker time dependence of $N(t-\tau)$ we replace the kernel $P_{i}(K,t-\tau)N(t-\tau)$ by the exponent $qe^{-\gamma(t-\tau)}$ where all time dependences are absorbed in  $\gamma$ and  all prefactors are absorbed in $q$. This deterministic continuous approximation  does not account for the time delay between the publication of the parent paper and the appearance of first citations, it cannot be used for quantitative estimates, and we use it here mostly for pedagogical reasons.

After all  substitutions Eq. \ref{model} reduces to
\begin{equation}
k_{i}(t)=\eta_{i} m_{dir}(t)+\int_{0}^{t}q_{i}ke_{i}^{-\gamma(t-\tau)} d\tau.
\label{SM-dynamics1}
\end{equation}
For a given $\eta_{i}$ it yields  $k_{i}(t)$. Here, $\eta_{i}$ is a number, specific for each paper (paper's fitness), $m_{dir}(t)$ is a known function, such as $\int_{0}^{\infty}m_{dir}(t)dt=1$, and $q_{i}=q(K_{i})$ is a known function of accumulated citations, $K_{i}=\int_{\tau=1}^{t}k_{i}(\tau)d\tau$. Equation \ref{nonlinearity} yields
\begin{equation}
q_{i}=a+b\ln K_{i}
\label{q}
\end{equation}
where for Physics papers published in 1984 $a=0.19,b=0.069,\gamma=1.2$ yr$^{-1}$, and  $m_{dir}(t)$ is shown in SM.

Equation \ref{SM-dynamics1} is an integral Volterra equation of second kind. Since its kernel $q_{i}$ depends on $K(t)$, this is a nonlinear equation and  its analytical solution is unknown. To find an approximate solution we note that $q_{i}(K_{i})$ dependence is logarithmic, namely, weak. Therefore, we can  consider $q_{i}$ as a parameter and solve  Eq. \ref{SM-dynamics1} for $q_{i}=const$. This yields
\begin{equation}
k(t)=\eta\left[m_{dir}(t)+q\int_{0}^{t}m_{dir}(\tau)e^{-(\gamma-q)(t-\tau)}d\tau\right].
\label{SM-dynamics2}
\end{equation}
where index $i$ has been omitted for brevity. Equation \ref{SM-dynamics2} indicates that each direct citation, captured by the term $\eta m_{dir}(t)$, induces a cascade of indirect citations that decays if  $q<\gamma$ and propagates if $q>\gamma$. The former case corresponds to ordinary papers, the latter case corresponds to runaways.

To find the total number of citations $K$ we integrate  Eq. \ref{SM-dynamics2}. This requires some analytical expression for $m_{dir}(t)$.  We crudely approximate it by the exponential dependence, $m_{dir}(t)=\delta e^{-\delta t}$, substitute it into Eq. \ref{SM-dynamics2}, and find

\begin{equation}
K(t)=K_{\infty}\left[1-\frac{(\gamma-\delta)(\gamma-q)e^{-\delta t}-q\delta e^{-(\gamma-q)t}}{\gamma(\gamma-q-\delta)}\right]
\label{SM-dynamics4}
\end{equation}
where
\begin{equation}
K_{\infty}=\frac{\eta\gamma}{\gamma-q}
\label{SM-dynamics5}
\end{equation}
For $q<\gamma$ Eq. \ref {SM-dynamics4} yields citation dynamics with saturation, $K\rightarrow K_{\infty}$, while for  $q>\gamma$ it predicts exponential growth, $K \sim e^{(q-\gamma)t}$.

The time constant accounting for approach to saturation (obsolescence rate $\Gamma$) is given by the smallest of the two exponents: $\delta$ and $\gamma-q$. For low-cited papers  $\Gamma\approx\delta$, for medium-cited papers   $\Gamma\approx\gamma-q$. We substitute Eq. \ref{q} into the latter relation and find
\begin{equation}
\Gamma=(\gamma-a)-b\ln{K}
\label{gamma_calc}
\end{equation}
We recover Eq. \ref{logarithm} where $\Gamma_{0}=\gamma-a$. Thus, different longevity of the highly-cited and ordinary papers is directly related to the nonlinear coefficient $b$ in Eq. \ref{q}.

In the framework of this  deterministic approach the citation career of the paper is set by its fitness $\eta$. Citation career of a low-fitness paper eventually comes to saturation, while citation career of a high-fitness paper may continue indefinitely. To find the critical fitness that marks the onset of the runaway behavior we consider the papers with relatively low fitness that do come to saturation. For these papers $K\rightarrow K_{\infty}$ in the long time limit. We revert Eq. \ref{SM-dynamics5} and find
\begin{equation}
\eta=K_{\infty}\left(1-\frac{q(K_{\infty})}{\gamma}\right)
\label{eta}
\end{equation}
Solution of this transcendental equation yields $K_{\infty}(\eta)$. This solution exists only for $\eta<\eta_{crit}$ where $\eta_{crit}$  is found from the condition $\frac{d K_{\infty}}{d\eta}=\infty$ (or $\frac{d\eta}{d K_{\infty}}=0$). We substitute   Eq. \ref{q} into Eq. \ref{eta}, perform differentiation, and find
\begin{equation}
\eta_{crit}=\frac{b}{\gamma} e^{\frac{\gamma-a}{b}-1}
\label{eta-crit}
\end{equation}
Equation \ref{eta-crit} shows that the fitness threshold for the runaway behavior sensitively depends on the nonlinear coefficient $b$. As expected, in the case of linear dynamics  when $b\rightarrow 0$, $\eta_{crit}\rightarrow\infty$, namely, the runaways disappear.

\section{Discussion}
The nonstationary nature of citation distributions has been already noticed. Redner \cite{Redner2005} found that citation lifetime of Physics papers increases with the number of citations. Lehmann, Jackson, and Lautrup \cite{Lehmann2005}  analyzed the  lifetime of high-energy physics papers, as defined through the fractions of the "live" and "dead" papers, and also found that the lifetime increases with the number of citations. Baumgartner and Leydesdorff  \cite{Baumgartner2013} showed that citation trajectories  of highly-cited papers are qualitatively different from the rest and do not come to saturation. The densification law established by Lescovec et al. \cite{Leskovec2007} states that, as time passes, the growing citation networks do not rest self-similar but  shrink in diameter and become denser. Our  observation of nonstationarity of citation distributions is in line with  these studies. However, we make a further step and find the explanation of this surprising fact.

First of all, there is a trivial source of nonstationarity- long citation lifetime. Indeed, our measurements  indicate that the characteristic citation lifetime of the ordinary Physics, Economics, and Mathematics papers is $\tau_{0}=4.6, 9$ and  11.8 yr, correspondingly (Fig. \ref{fig:lifetime}). Thus, if we consider citation distributions in the time window of, say, 15 years, the distributions for ordinary Physics papers  should be already stationary, while those for Economics and Mathematics are not. Many  citation distributions were measured in the time window of only 5-10 years after publication, when even ordinary papers didn't achieve saturation, hence  it is not a surprise that such distributions exhibited different shapes.

A more important source of nonstationarity is the fact that citation lifetime of a paper increases with the number of citations, and this seems to hold  for all research fields. When this number exceeds a certain  critical number $K_{0}$, specific for each field and publication year, citation lifetime goes to infinity, in such a way that the papers with $K>K_{0}$ become runaways or supercritical papers.\cite{Bianconi2001}  We explain the origin of the runaway behavior  within the framework of the copying mechanism. Indeed, every newly published paper induces a slowly decaying train of  citing papers whose authors find it in databases, Internet, etc. Each of these citing papers induces a cascade of secondary citing papers whose authors can copy the source paper into their reference list. Whether this cascade decays or propagates- this depends on the reproductive number $R=PN$, where $N$ is the average number of the second-generation citing papers per one first-generation citing paper (the fan-out coefficient), and $P$ is the probability of  indirect citation (copying). For ordinary papers $R<1$ and the citation  cascade decays, for highly-cited papers $R>1$, and  the citation cascade propagates indefinitely in time. The papers  that managed to overcome the tipping point of  $R=1$ are  runaways and their presence can result in a "winner-takes-all" network.\cite{Krapivsky2001,Vazquez2003}

What are the implications of  nonstationarity of citation network  for its degree distribution?  If we focus on the set of papers in one field published in one year, its citation distribution at early times mimics the fitness distribution, while at later times it develops according to our model (Eq. \ref{model}). While initial citation distribution is close to log-normal, the evolving distributions straighten  and become closer to power-law. This corroborates the  conjecture of Mokryn and Reznik \cite{Mokryn2015} who assumed that the power-law degree distribution can be found only in static networks, namely, those that underwent a long period of development. We show here that citation distributions,  once considered as static, are in fact transient. As time passes, they evolve from the convex shape  to straight line, and then some of them can even acquire the concave shape in the log-log plot. The rate of this evolution is field-specific and depends on fitness distribution, on the growth of the number of publications  in this field, and on the average reference list length.

So, what is the source of the "power-law" citation distributions? Our findings indicate that these are inherited from the fitness distribution and then modified during citation process. The question of why citation distributions follow the power-law dependence is thus relegated to fitness distribution. The deep question is why fitness distribution  has  certain shape? We found that the fitness distribution for Physics papers published in 1984  can be fitted by a log-normal distribution with $\sigma\sim 1.1$. Interestingly, we found that the fitness distributions for Mathematics and Economics papers published in 1984  follow the log-normal distribution  with the same $\sigma$ (but different $\mu$). We leave for further studies this issue of universality of fitness distributions and only mention in passing that the log-normal distribution with $\sigma\sim 1.1$  is one of the narrowest log-normal distributions observed in science.\cite{Limpert2001}


Does power-law degree distribution necessarily imply a scale-free network? The notion of thscale-free networks was introduced by Barabasi and Albert  \cite{AlbertBarabasi2002} who realized that the ubiquity of the power-law degree distributions in complex  networks  implies some universal generating mechanism. The Barabasi-Albert preferential attachment mechanism  generates a growing complex network whose degree distribution becomes stationary in the long time limit. This stationary shape is the power-law degree distribution  with the universal exponent $\alpha=3$ and this implies the scale-free network.  Before the advent of complex networks, the  scale-free phenomena have been usually associated with phase transitions and critical point in condensed matter physics. The characteristic scale- correlation length- diverges at critical point and this results in universal power laws for static and dynamic properties of the substance. The universal power-law degree distribution in  complex networks has a great appeal to physicists with their quest for grand unified theories since it implies that the properties of such diverse objects- complex networks   and substances at critical point- are described by the same mathematical formalism.  However, the equivalence between the power-law degree distribution and the scale-free character of a complex network holds only if the degree distribution is stationary or at least does not change its shape.   However, for nonstationary networks there  can be a dynamic scale that governs the network development. We demonstrate here that citation distributions are nonstationary, their shape changes with time, and there is a certain dynamic scale that is important for network growth. Therefore, the scale-free character of citation networks is limited to its degree distribution at each temporal snapshot.

Can our findings  be extended to other complex networks? We found that nonstationary citation distribution is related to nonlinear citation dynamics. Citation dynamics of patents is also nonlinear.\cite{Zhou2004,Csardi2007,Higham2017} Thus, we expect that patent citation distribution is  nonstationary as well.  With respect to degree distribution in  Internet networks- these networks have an important distinction:  unlike citation networks which are directed and acyclic, the WWW is not and the links there can be edited.  Nevertheless,  runaways   were detected in the distribution of the Web page popularity as well.\cite{Kong2008}

In summary, we have demonstrated that while statistical distribution of citations to scientific papers can be fitted by the  power-law dependence, this distribution is nonstationary and does not acquire a limiting  shape in the long time limit. While the power-law fit implies the scale-free citation network, we show that  there is a hidden dynamic scale associated with the onset of runaways. Thus, the similarity of measured citation distributions to the power-law dependence is superficial and does not have deep implications.

\begin{acknowledgments}
I am grateful to Sorin Solomon and Peter Richmond who introduced me into the fascinating field of the power-law distributions.  
\end{acknowledgments}
\bibliography{references_master}
\section{Supplementary Material}
\section*{Model parameters}
\begin{figure}[!ht]
\begin{center}
\includegraphics*[width=0.45\textwidth]{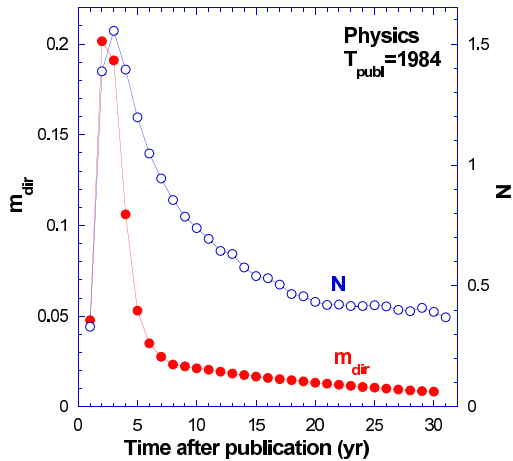}
\caption{Time dependence of  $m_{dir}(t)$, the reduced number of direct citations, and  $N(t)$, the mean number of second-generation citing papers per one first-generation citing paper.
}
\label{fig:m_dir-N}
\end{center}
\end{figure}

Our model relies on the empirical functions $m_{dir}(t)$ and $N(t)$.  The former is the normalized  rate of direct citations and it is related to our definition of fitness $\eta$. Indeed, we define $\eta$ from the relation $k_{dir}=\eta m_{dir}$ where $k_{dir}$ is the direct citation rate of a paper and  $\int_{0}^{t}m_{dir}(\tau)d\tau=1$ when $t\rightarrow \infty$ and the publication year corresponds to $t=1$. If the integral converges very slowly we just set $t=25$ yr. Figure \ref{fig:m_dir-N} shows $m_{dir}(t)$ found  by analyzing dynamics of direct and indirect citations of 37 Physical Review B papers published in 1984.\cite{Golosovsky2017} This function sharply increases, achieves its maximum 1-2 years after publication and then slowly decreases. Its time dependence is related to the age composition of the average reference list  and on the growth of the number of publications. In the long time limit these two factors have opposite trends and that is why $m_{dir}$ decays so slow. The function $m_{dir}(t)$ is specific for each discipline. There is an indication that for the Mathematics and Economics papers  $m_{dir}(t)$ does not go to zero in the long time limit and may even remain stationary. Indeed, the growth of the number of publications in these fields is faster than for the Physics area, while the tendency to cite old papers is  stronger. Both these factors conspire in that the mean number of citations for these fields does not decay with time (Fig. 4 of the paper).

The function $N(t)$ is defined as follows. For each source paper we measured the number of the first-generation citations. Obviously, this number is equal to the number of first-generation citing papers. For each of the latter we measured the average number of citations $M^{nn}$ and citing papers $N^{nn}$. These are second-generation citations and citing papers and their numbers are not equal since one second-generation citing paper can cite several first-generation citing papers. In fact, $M^{nn}$ is well-known in network science- this is an average number of links to next-nearest neighbors. We measured  $M^{nn}$ and $N^{nn}$ for several sets of Physics papers and found that $N^{nn}=0.83M^{nn}$ in the long time limit.\cite{Golosovsky2017}  We didn't measure time dependence of $M^{nn}$ and $N^{nn}$ but assumed that the time dependence of $M^{nn}(t)$ is the same as that of $M(t)$- the average number of the first-generation citations. Figure \ref{fig:m_dir-N} plots $N^{nn}(t)$ found from the above considerations. During first 1-3 years after publication $N(t)$  mimics $m_{dir}(t)$ since most part of citations are direct. Later on, these dependences become different due to appearance of indirect citations. It shall be noted, however, that our numerical simulations strongly depend on $m_{dir}(t)$ at all times and on $N(t)$ during first 2-3 years, hence the tail of $N(t)$ in this context is irrelevant.
\section*{Concave citation distribution}
\begin{figure}[!ht]
\begin{center}
\includegraphics*[width=0.48\textwidth]{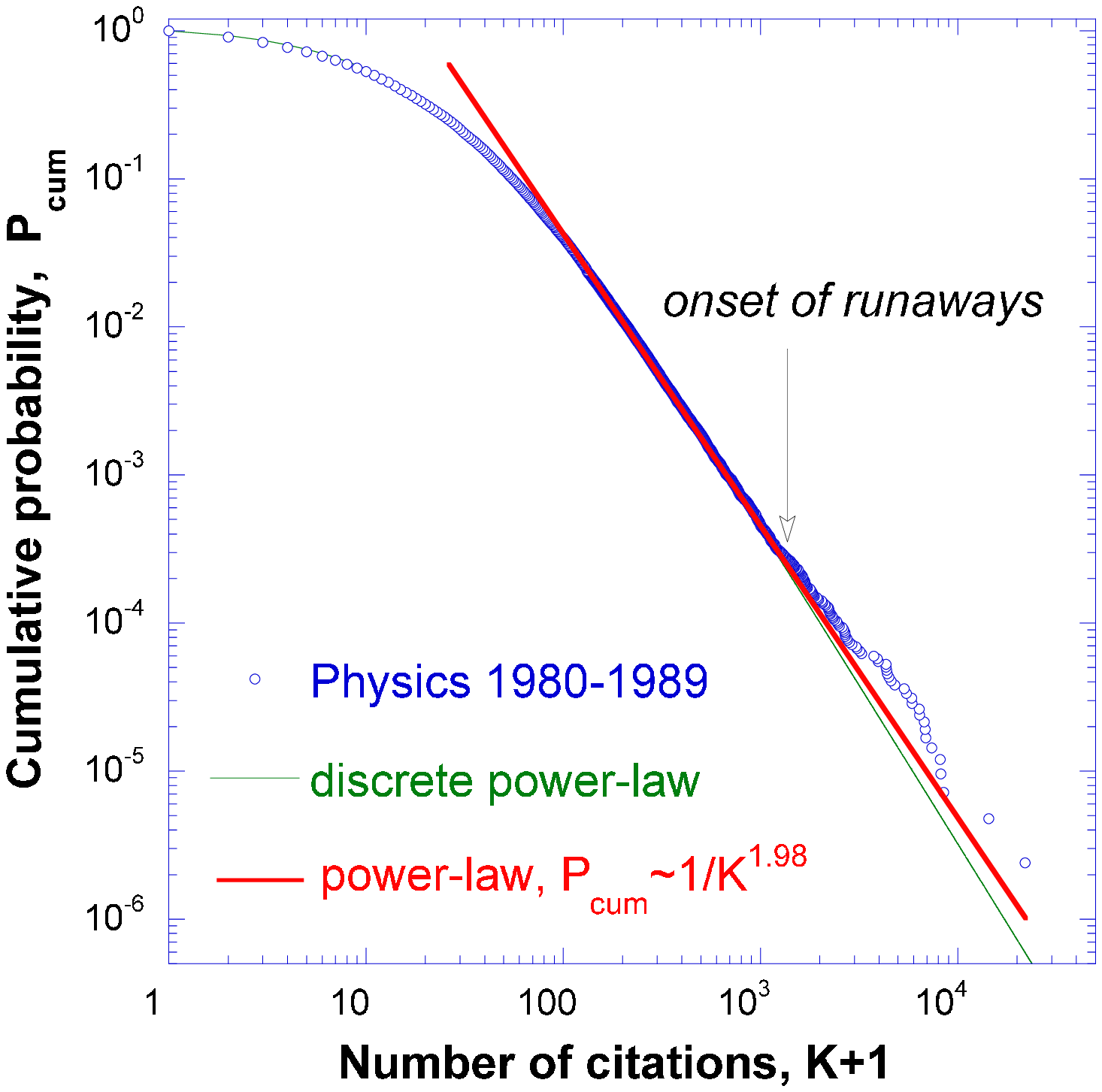}
\includegraphics*[width=0.48\textwidth]{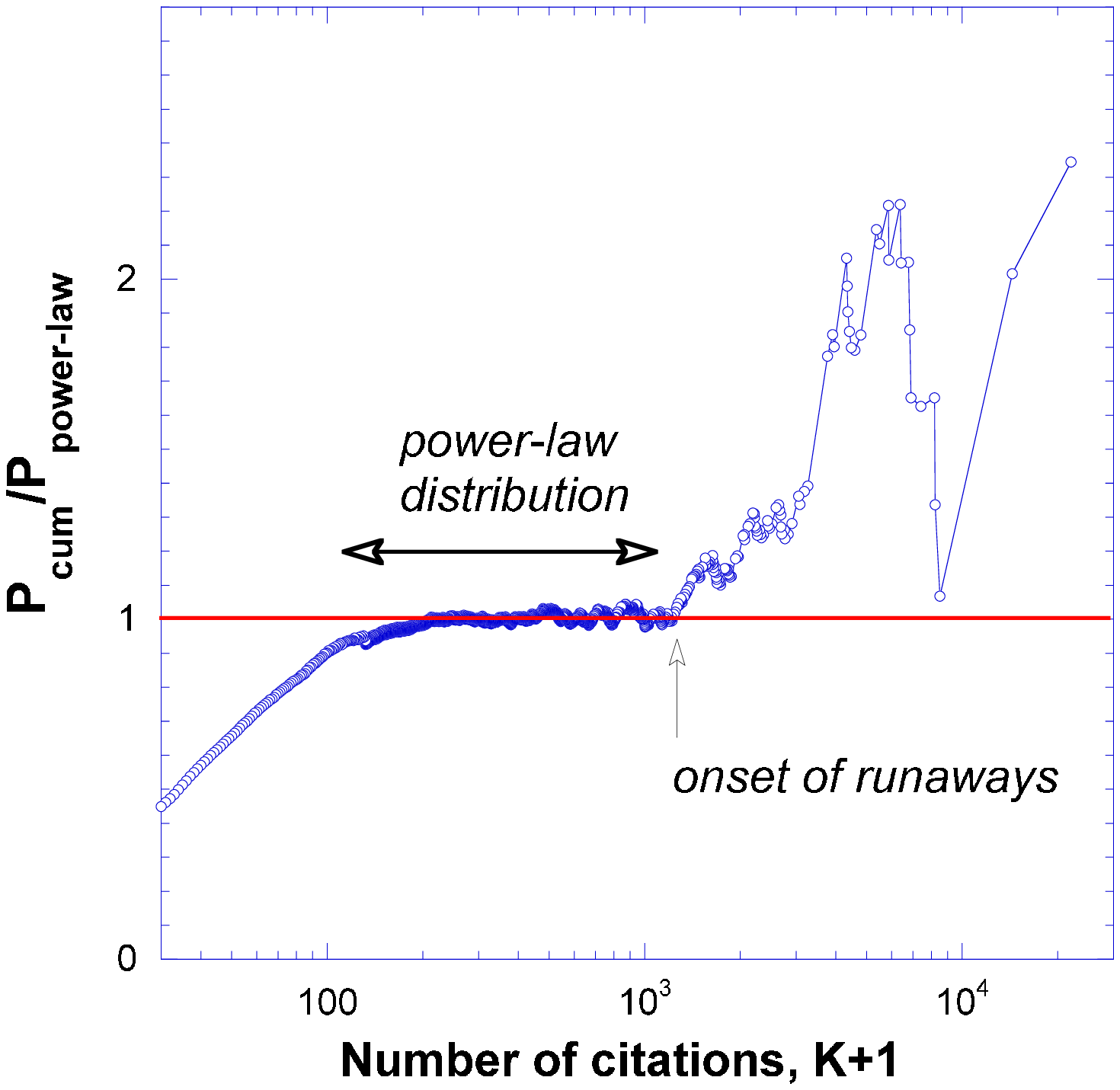}
\caption{Left panel: citation distribution for 418,438 Physics papers published in 1980-1989 (overviews excluded) measured in the year 2008 (open blue circles).\cite{Golosovsky2012a} The green continuous curve shows discrete-power-law (Waring) distribution. It accounts fairly well for  $0\leq K<1200$ and fails for the tail, $K>1200$.  The red straight line shows a power-law fit, $P_{cum}\propto K^{-1.98}$. This fit accounts for the body of the distribution, $100<K<1200$, and fails for the tail. Note upward deviation of the tail of the measured distribution from both fits, namely, concave cumulative distribution. Right panel: the ratio of the measured cumulative distribution to the power-law fit. While the deviation for $K<100$ is trivial (it is related to the rounding off of the distribution at small $K$), the upward deviation at $K>1200$  is not trivial and indicates on the runaway tail. This tail contains  130 papers.
}
\label{fig:runaways}
\end{center}
\end{figure}

\begin{figure}[!ht]
\begin{center}
\includegraphics*[width=0.38\textwidth]{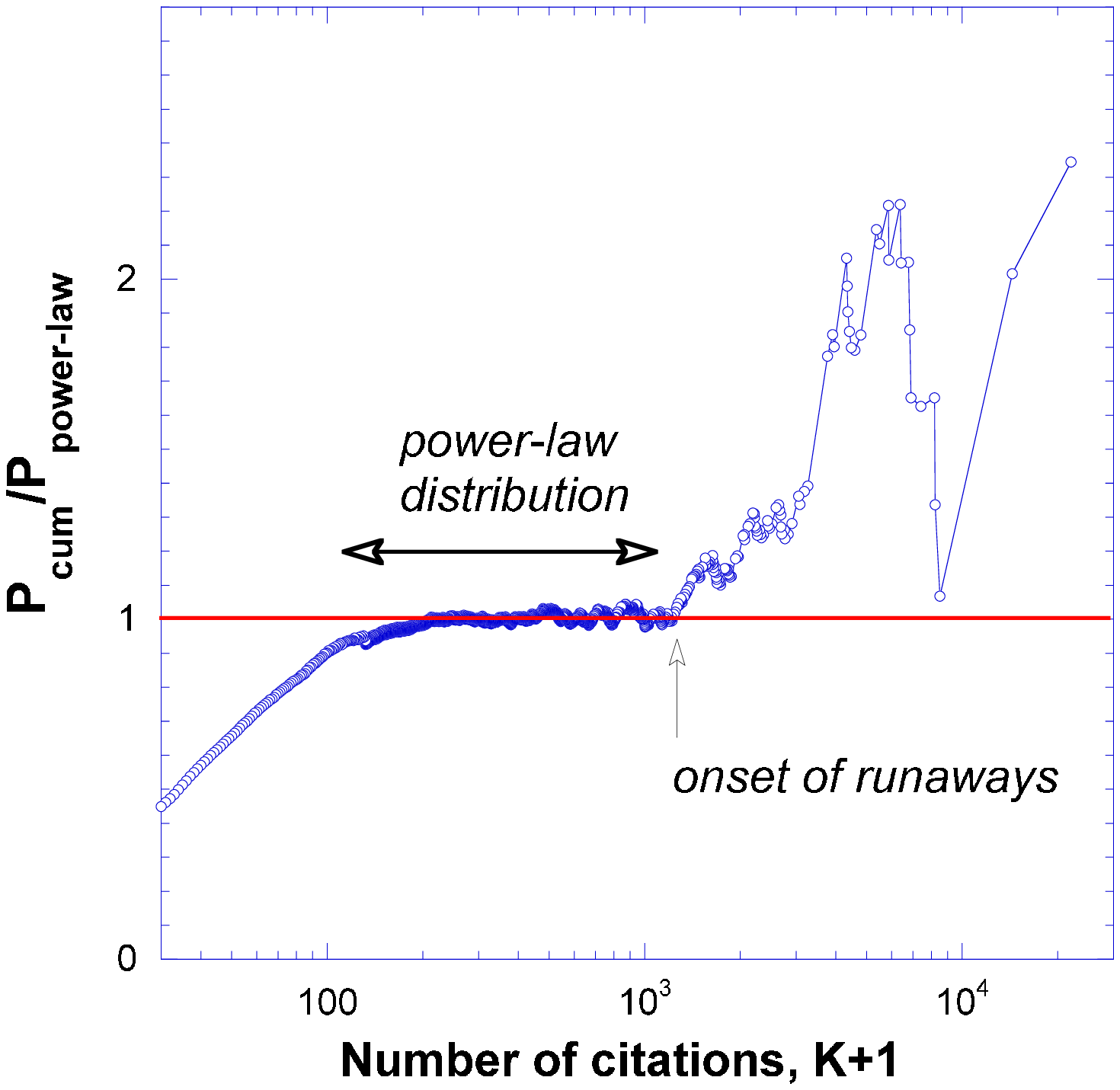}
\caption{The ratio of the measured cumulative distribution to the power-law fit. While the deviation for $K<100$ is trivial and is related to the rounding off of the distribution at small $K$, the upward deviation at $K>1218$  is not trivial and indicates on the runaway tail. This tail contains  130 papers.
}
\label{fig:ratio}
\end{center}
\end{figure}

\end{document}